\documentclass{pasj00}

\begin{document}
\SetRunningHead{Y. Terada et al.}{Suzaku X-ray Observation of BD+43$^\circ$3654}
\Received{2012/06/05}
\Accepted{2012/07/23}

\title{Search for Diffuse X-rays from the Bow Shock Region of Runaway Star BD+43$^\circ$3654 with Suzaku}

 \author{%
  Yukikatsu \textsc{Terada} \altaffilmark{1}
  Makoto S. \textsc{Tashiro} \altaffilmark{1}
  Aya \textsc{Bamba} \altaffilmark{2}
  Ryo \textsc{Yamazaki} \altaffilmark{2}
  Tomomi \textsc{Kouzu} \altaffilmark{1}
  Shu \textsc{Koyama} \altaffilmark{1}
  and
  Hiromi \textsc{Seta} \altaffilmark{3}}
  \altaffiltext{1}{Graduate School of Science and Engineering, Saitama University, 255 Simo-Ohkubo, Sakura-ku, Saitama City, Saitama 338-8570, Japan}
  \email{terada@phy.saitama-u.ac.jp}
  \altaffiltext{2}{Department of Physics and Mathematics, College of Science and Engineering, Aoyama Gakuin University, 5-10-1 Fuchinobe, Chuo-ku, Sagamihara, Kanagawa 252-5258, Japan}
  \altaffiltext{3}{Department of Physics, Science, Rikkyo University, 3-34-1, Nishi-Ikebukuro, Toshima-ku, Tokyo, Japan}


\maketitle

\begin{abstract}
The bow shocks of runaway stars with strong stellar winds of over 
2000 km s$^{-1}$ can serve as particle acceleration sites. 
The conversion from stellar wind luminosity
into particle acceleration power has an efficiency of
the same order of magnitude as those in supernova remnants, 
based on the radio emission from the bow shock region 
of runaway star BD+43$^\circ$3654 \citep{Benaglia10}.
If this object exhibits typical characteristics, 
then runaway star systems can contribute a non-negligible fraction
of Galactic cosmic-ray electrons.
To constrain the maximum energy of accelerated particles 
from measurements of possible non-thermal emissions in the X-ray band, 
Suzaku observed BD+43$^\circ$3654 in April 2011 with an exposure of 99 ks.
Because the onboard instruments have a stable and low background level,
Suzaku detected a possible enhancement over the background 
of $7.6\pm 3.4$ cnt arcmin$^{-2}$ at the bow shock region,
where the error represents the 3 sigma statistics only.
However, the excess is not significant within the systematic errors of 
non-X-ray and cosmic-ray backgrounds of the X-ray Imaging Spectrometer,
which are $\pm 6.0$ and $\pm 34$ cnt arcmin$^{-2}$, respectively,
and the 3-sigma upper limit in the X-ray luminosity 
from the shock region, which is $1.1 \times 10^{32}$ erg s$^{-1}$ per 41.2 arcmin$^2$
in the 0.5 to 10 keV band.
This result leads to three conclusions:
(1) a shock-heating process is inefficient on this system;
(2) the maximum energy of electrons does not exceed $\sim$ 10 TeV, 
corresponding to a Lorentz factor of less than $10^7$; and
(3) the magnetic field in the shock acceleration site might not be as turbulent
as those in pulsar wind nebulae and supernova remnants. 
\end{abstract}

\section{Introduction}
\label{section:introduction}
\subsection{Runaway stars as particle acceleration sites}
\label{section:introduction:acceleration}
An important goal of astrophysics is to understand 
the non-thermal phenomena in the universe. 
In our galaxy, pulsars and X-ray binaries are well-known objects 
that generate high-energy electrons and possibly protons, 
which emit non-thermal photons 
(e.g., \cite{Sturrock71,Ruderman75,Cheng86a,Cheng86b,Harding90}).
Recently, sensitive searches of non-thermal X-rays have led to 
magnetic white dwarfs also being considered as possible 
non-thermal emitters \citep{terada08b,terada10,terada11}. 
In addition to rotating magnetospheres, like pulsars, 
shocks in supernova remnants (SNRs) and pulsar wind nebulae (PWNe)
are the most common Galactic cosmic-ray acceleration sites
(e.g., \cite{Shklovskii53,Ginzburg53,Bell78,Lagage83,kennel84,koyama95,Rees74}).
Searching for new types of particle-acceleration sites is 
a possible key to understanding the non-thermal universe. 

In this paper, we focus on early-type massive stars with 
velocities greater than $30$ km s$^{-1}$, that is, runaway OB stars, 
as another source of non-thermal emission.
The most prominent characteristics of runaway OB stars are 
their high spatial velocities 
and their very fast stellar winds. 
The velocities of the stellar winds 
reach a few thousand kilometers per second, which 
is comparable with 
the shock speeds of young SNRs 
(\cite{Bamba05} and references therein).
Thus, in the rest frame of an OB star,
two shocks (a bow shock and a wind termination shock)
and a contact discontinuity appear in order
to balance the ram pressure of the interstellar gas
with that of the stellar wind \citep{delValle12}.
Bow shock structures can often be seen
in infrared images of runaway OB stars \citep{Buren88,peri12}.
\citet{Benaglia10} have reported radio observations in the NRAO-VLA Sky Survey program, indicating the presence of high-energy electrons in the bow shock region of runaway star BD+43$^\circ$3654. 

The object BD+43$^\circ$3654 is thought to be a blue straggler system formed by a close encounter between two tight massive binaries \citep{Gvaramadze08} in the core of the Cygnus OB2 association (\cite{Comeron02} and references therein),
which is the most massive OB association in the solar neighborhood,
at a distance of $d = 1.4$ kpc \citep{Hanson03,Comeron07}.
The spectral type of the star is O4If with an age of about 1.6 Myr 
and a stellar mass of $70 \pm 15 ~M_\odot$ \citep{Comeron07};
the star is one of the most massive runaway stars known today.
The heliocentric radial velocity is measured to be 
$v_\star = -66.2 \pm 9.4$ km s$^{-1}$ \citep{Kobulnicky10}. 
The typical mass-loss rate for O4I stars with a stellar wind velocity of $v_{\rm sw} = $ 2300 km~s$^{-1}$ is $\dot{M} = 10^{-5} M_\odot $ yr$^{-1}$
\citep{Markova04,Repolust04}.
The bow shock structure was found by IRAS observations in a systematic 60 $\mu$m survey \citep{Buren88}, 
and was confirmed by the Midcourse Space eXperiment 
in the D and E bands \citep{Comeron07}. 
Synchrotron emission was also found from this region in the radio bands at 1.42 GHz and 4.86 GHz, as already mentioned \citep{Benaglia10}.

\subsection{Similarity with SNR cases}
\label{section:introduction:SNRcomparison}
We can investigate the runaway system 
as a shock acceleration site by using its similarities 
with SNRs: velocity of stellar wind, 
system size, environment of inter-stellar matter, etc.
If we adopt the typical parameters of stellar winds from O4I stars 
\citep{Markova04,Repolust04},
the stellar wind has a kinetic luminosity up to
\begin{eqnarray}
\dot{E}_{\rm sw} &=& 2 \times 10^{37} 
\left(\frac{\dot{M}}{10^{-5}M_{\odot} ~{\rm yr}^{-1}}\right) 
\nonumber \\ 
&&
\times \left(\frac{ v_{\rm sw}}{2300 ~{\rm km~s}^{-1}}\right)^2
~{\rm erg ~s}^{-1}.
\label{eq:momentum_sw}
\end{eqnarray}
The termination shock causes particle acceleration 
which generates high-energy protons and electrons \citep{delValle12}.
In order to account for the radio luminosity of BD+43$^\circ$3654
reported by \citet{Benaglia10},
high energy electrons with the energy index of $p \sim 2.0$
should have a total energy of $E_{\rm e} \sim  2 \times 10^{45}$ ergs,
under the assumption of equipartition between the energy densities of 
electrons $u_{\rm e}$ and magnetic field $u_{\rm B}$,
as is the case with many young SNRs \citep{bamba03, Bamba05}. 
In this case, corresponding magnetic-field strength $B$ is about
a few tens of micro Gauss.
Then, the cooling timescale of electrons emitting
synchrotron radiation with characteristic frequency $\nu$ \citep{Reynolds99},
\begin{equation}
\tau_{\rm syc} \sim 
10^{15} \left(\frac{B}{10 ~\mu{\rm G}}\right)^{-3/2} 
\left(\frac{\nu}{1 {\rm ~GHz}}\right)^{-1/2}
~{\rm s},
\label{eq:timescale_sync}
\end{equation}
is much longer than the dynamical timescale,
\begin{equation}
\tau_{\rm dy} \sim \frac{R_0}{v_{\rm sw}} = 3 \times 10^{10} ~{\rm s},
\label{eq:timescale_dy}
\end{equation}
where $R_0 \sim 2$ pc is the distance to the shock front 
of the bow shock measured by \citet{Benaglia10}.

Electrons are accelerated and emitting synchrotron radiation 
around the shock region. 
They escape from the shock region flowing out along the stream 
backward of the star in a certain time scale. 
Assuming that the system is in a steady state, the electron 
production rate $L_{\rm e}$ is roughly identical to the loss rate 
from the system. 
In that case, $L_{\rm e}$ will be described by $E_{\rm e}$ 
divided by the escaping time scale. 
Here, we regard the dynamical effects should be dominant, 
since the cooling time scale of the electrons is fairly longer 
than the dynamical time scale, $\tau_{\rm dy}$, 
as described above (equations (\ref{eq:timescale_sync}) 
and (\ref{eq:timescale_dy})).
Thus a rough estimate of $L_{\rm e}$
is $\sim E_{\rm e}/\tau_{\rm dy} = 6 \times 10^{34}$ erg s$^{-1}$, 
which corresponds to about 0.3\% of $\dot{E}_{\rm sw}$ 
given in equation (\ref{eq:momentum_sw}).
Therefore, the conversion from the energies of stellar winds 
into the acceleration power of particles 
in this runaway star system has an efficiency of
the same order of magnitude as those in SNRs.

Since about 10\% of O-stars exhibit bow shock structures
\citep{peri12}, the $\sim2,000$ runaway stars in our Galaxy
can produce a non-negligible proportion of about 10\% 
of the cosmic-ray electrons observed at Earth 
(i.e., $10^{38}$ erg s$^{-1}$),
if BD+43$^\circ$3654, with its high-energy electron production rate of $6 \times 10^{34}$ erg s$^{-1}$, represents the typical case of 
a runaway-star system as a particle acceleration site.
Therefore, it is important to study physical parameters 
of the particle acceleration process in BD+43$^\circ$3654.

In the sense of energetics, BD+43$^\circ$3654 is an SNR-equivalent 
particle acceleration site. Then, the next studies should be 
how much energy the particles gain in the system.
The maximum energy of accelerated electrons, $E_{\rm max}$,
is measured by the roll-off frequency, $\nu_{\rm roll}$,
of the synchrotron radiation. 
If $E_{\rm max}$ is $\sim$ 8--20 TeV as suggested by \citet{Benaglia10},
then, according to equation (2) in \citet{Reynolds99}, 
$\nu_{\rm roll}$ should be $ \sim $$10^{16}$--$10^{20}$ Hz, 
which comes into the X-ray band.
If the spectrum were simply extended from the radio band 
so that $\nu_{\rm roll}=\infty$, then
X-rays should be clearly detected with the sensitivities 
of current missions. In that case the absence of X-rays would 
provide a unique upper limit $E_{\rm max}$.
However, high sensitivity to diffuse X-ray emissions 
is required in this case.
In this study, we performed an X-ray search for 
possible synchrotron emission from 
the shock region of BD+43$^\circ$3654
with the X-ray satellite Suzaku \citep{Mitsuda07}.

The rest of this paper is organized as follows. We describe the X-ray observations
of this object in Section \ref{section:obs}, 
summarize the results of Suzaku analyses in Section \ref{section:ana}, 
and discuss our results in Section \ref{section:discussion}.

\section{Observation and Data Reduction}
\label{section:obs}
\subsection{Suzaku observation of BD+43$^\circ$3654}
\label{section:obs_obs}
{\it Suzaku} is the fifth in a series of Japanese X-ray satellites 
having sensitivities in the 0.2--600 keV band \citep{Mitsuda07}.
It carries two X-ray instruments on board: 
the X-ray Imaging Spectrometer (XIS) 
for the 0.2--12 keV band \citep{xis2007}, 
and the Hard X-ray Detector (HXD) 
for the 12--600 keV band \citep{hxd2007a}.
Although both instruments have very low-background signals \citep{xis_bgd08, hxd_bgd09}, which enable us to perform high sensitivity surveys of X-rays, the Suzaku XIS is most suitable for our purpose because it is capable of imaging and stable backgrounds.

We observed BD+43$^\circ$3654 and the bow shock region with Suzaku 
from 2011 April 4 1:53~UT to April 6 2:13~UT (OBSID=506004010).
To put the bow shock region at the XIS nominal pointing position, 
the aiming point was set to ($\alpha, \delta$)[J2000] 
= ($20^{\rm h}33^{\rm m}40.00^{\rm s}$, $+44^{\circ}03^{\prime}00.0^{"}$),
although the runaway star is at ($\alpha, \delta$)[J2000] 
= ($20^{\rm h}33^{\rm m}36.077^{\rm s}$, $+43^{\circ}59^{\prime}07.40^{"}$).
The XIS was operated in the normal clocking mode, 
without window/burst options 
but with the Space Charge-Injection (SCI) function \citep{xis_sci08}.
The XIS consists of three front-side illuminated (FI) CCD chips
(XIS0, XIS2, and XIS3) and one back-side illuminated (BI) chip (XIS1).
XIS2 was not operational due 
to damage from a meteoroid strike.
HXD was operated in nominal mode: 
half of the 64 PIN diodes were operated with a voltage of 400 V 
and the others at 500 V, and the PMTs were operated at nominal gain.

\subsection{Data reduction}
\label{section:obs_reduction}
We used the observation datasets processed by the standard 
Suzaku pipeline version 2.5.16.29, 
with the calibration version (CALDBVER) of 
hxd20101202, xis20101108, xrt20100730, and xrs20060410.
We used \emph{ftools} from the HEADAS 6.11 package 
with XSPEC version 12.7.0.

The source was detected by the XIS in the 0.5--10 keV band
at an average count rate of 0.15 and 0.28 cnt s$^{-1}$ per sensor 
in the FI- and BI- CCD cameras, respectively.
Cleaned events of the XIS data were obtained 
using the standard criteria of the pipeline process.
The total exposure of the XIS was 99.0 ks.

We did not use the HXD-GSO data 
because of sensitivity limitations.
According to the INTEGRAL catalog, there were no 
contaminating sources in the field of view (FOV) of 
the HXD-PIN detector.
The cleaned events from HXD-PIN were obtained 
by using the standard criteria of the process.
The total exposure of HXD-PIN was 82.5 ks.
The averaged count rate of the cleaned PIN data 
including backgrounds was 0.40 cnt s$^{-1}$ in the 13--70 keV band.
The non-X-ray background (NXB) events were estimated and 
provided by the HXD team. 
We used them with METHOD=`LCFITDT(bgd\_d)' and 
the METHODV=`2.0ver0804'.  
The average count rate of the NXB was 0.38 cnt s$^{-1}$ 
in the same energy range, 
and the residual count rate is consistent with the level 
of the cosmic X-ray background (CXB, 0.02 cnt s$^{-1}$) 
within the systematic errors of NXB \citep{hxd_bgd09}.
Therefore, HXD-PIN detected no significant signals 
in the 13--70 keV band, and thus 
we do not use the HXD data in the following analyses.

\section{Analyses and Results}
\label{section:ana}
\subsection{X-ray sources in the field of view}
\label{section:ana_image}

\begin{figure}[htb]
\centerline{\includegraphics[angle=0,width=0.45\textwidth]{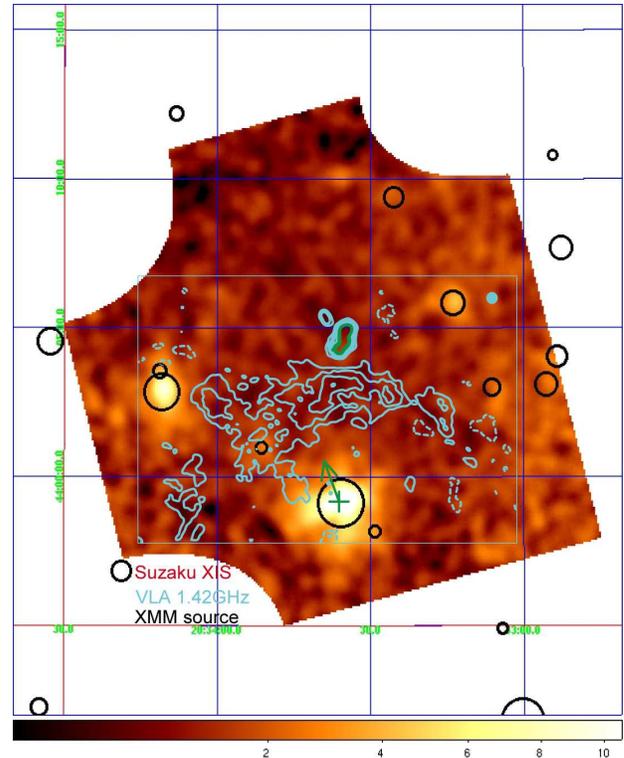}}
\caption{
The color map shows the X-ray image of BD+43$^\circ$3654 with the Suzaku XIS 
in the 0.5--10 keV band. 
The areas irradiated by the calibration sources of $^{55}$Fe 
are excluded from the image. 
The radio image in the 1.42 GHz band obtained by the VLA observation
\citep{Benaglia10} is overlaid as the contour image in cyan; 
the radio data is valid in the cyan box.
The black circles represent point sources 
detected by {\it XMM-Newton},
which are listed in Table \ref{tbl:xmm_points}.
The position and runaway direction of BD+43$^\circ$3654 are also shown 
in green with the cross and arrow, respectively.}
\label{fig:xray_image}
\end{figure}

We first checked the XIS images
in the 0.5--10 keV band.
The spatial distributions of the non X-ray backgrounds (NXBs) 
on the XIS chips in the 0.5--10 keV band
were estimated by the Suzaku \emph{ftool} {\it xisnxbgen} \citep{xis_bgd08} 
using the background database in the CALDB area 
for the period from 120 days before to 30 days after the observation. 
After the subtraction of the NXB image from the sky image with 
the XIS in the same energy bandpass, 
the vignetting effects of the X-ray telescopes (XRTs) were 
corrected by a simulated map calculated for a uniform sky input with
the ray-tracing tool {\it xissim} \citep{Ishisaki07}. 
In this simulation, we assume that the energy spectrum of 
incident photons follows the power-law shape with a photon index 
of 1.0. Changes of $\pm 1.0$ in the photon index correspond to 
changes in image shapes at the $\pm 5$\% level. 
The vignetting-corrected X-ray image from the XIS 
in the 0.5--10 keV band is shown in Fig.\ \ref{fig:xray_image}. 

It is important in diffuse X-ray observations to extract 
(or to estimate the flux of) dim point sources in the field of view, 
to reduce systematic errors caused by fluctuations of unresolved sources 
in CXB.
The brightest source is BD+43$^\circ$3654 itself;
the X-ray image around the object was consistent
with the point-spread function of the XRTs.
In the X-ray light curve, after barycentric correction 
by {\it aebarycen} \citep{terada08a}, 
we found neither significant flares 
nor coherent-periodic signals in the 16 to 10,000 s range.
The X-ray spectrum was well described by 
a MEKAL model \citep{mewe85, liedahl95, spex}
with a temperature of $0.60_{-0.09}^{+0.14}$ keV and 
a metal abundance of $0.32_{-0.17}^{+0.40}$ solar,
where the photoabsorption of the hydrogen column density was
$N_{\rm H} = 1.45_{-0.17}^{+0.26} \times 10^{22}$cm$^{2}$.
The emission measure and X-ray luminosity in the 0.5--10 keV band
were EM = $6.5_{-2.8}^{+9.9} \times 10^{55}$ cm$^{-3}$  
and $3.7_{-1.6}^{+5.6} \times 10^{31}$ erg s$^{-1}$, respectively.
The value of $N_{\rm H}$ is consistent with the galactic neutral hydrogen
map by \citet{lab}, and other parameters were well understood as being 
typical of X-ray emissions from O-type stars \citep{Naze11},
following the rough relation between $kT$ (keV) and EM (cm$^{-3}$) 
given by $\log {\rm EM} \sim 53.9 - \log kT$.

For the further rejection of contaminating sources, we checked
the archive data of BD+43$^\circ$3654 from 
{\it XMM-Newton} (OBSID = 0653690101, PI= Zabalza Victor, 
starting from 2010 May 8 08:04~UT, 46.7 ks exposure). 
Several faint point sources that contaminate the FOV of the XIS 
are listed in table \ref{tbl:xmm_points} 
and are plotted as black circles in Fig.\ \ref{fig:xray_image}.
Therefore, in the following analyses, we can exclude contamination 
by point sources with X-ray fluxes above 
$S_{\rm c} = 3.7 \times 10^{-14}$ erg s$^{-1}$ cm$^{-2}$
in the 0.5--10 keV band.

\begin{table}[h]
\begin{center}
\caption{Point sources detected with EPIC during the {\it XMM-Newton} 
observation of OBSID= 0653690101 (except for BD+43$^\circ$3654). 
The X-ray flux is in units of $10^{-13}$ erg s$^{-1}$ cm$^{-2}$ 
in the 0.5 to 10 keV band.
}
\begin{tabular}{cll}
\hline
RA & DEC & flux\\
\hline
308.5455 & 44.047592 & 1.2\\
308.54706 & 44.058725 & 0.47\\
308.23159 & 44.051389 & 0.77\\
308.5781 & 43.947109 & 0.63\\
308.22298 & 44.067196 & 0.70\\
308.3077 & 44.097479 & 0.83\\
308.27586 & 44.050153 & 0.53\\
308.37145 & 43.969254 & 0.40\\
308.3562 & 44.156242 & 0.67\\
308.46396 & 44.016647 & 0.37\\
\hline
\label{tbl:xmm_points}
\end{tabular}
\end{center}
\end{table}

\subsection{Upper limit of the X-ray flux from the bow shock region}
\label{section:ana_flux_diffuse}
\begin{figure}[htb]
\centerline{
\includegraphics[angle=0,width=0.45\textwidth]{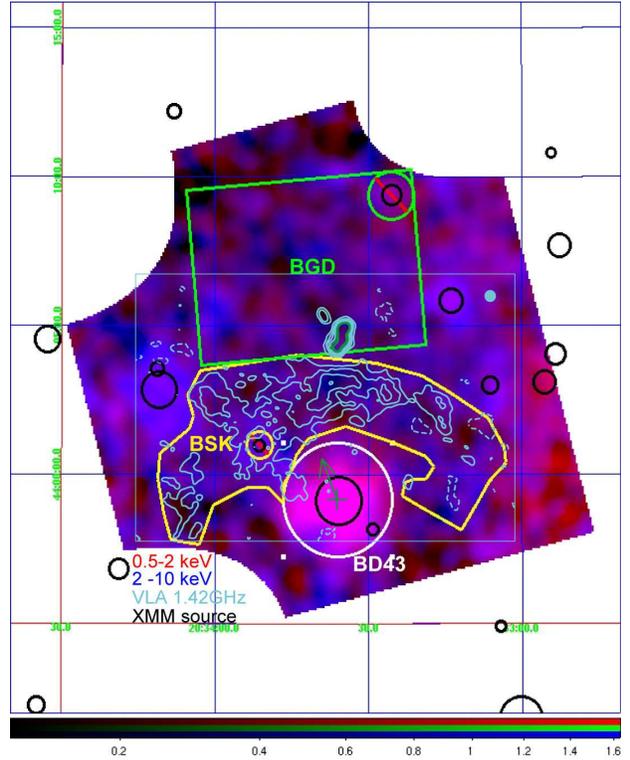}
}
\caption{
The energy resolved X-ray images from the Suzaku XIS.
The 0.5--2 keV and 2--10 keV images are shown in
red and blue, respectively. 
Point sources from {\it XMM-Newton} are plotted as black circles, 
and the radio contour in the 1.42 GHz band \citep{Benaglia10} is 
displayed in cyan.
The definitions of regions, BGD, BSK, and BD43, are outlined
in green, yellow, and white, respectively.}
\label{fig:xray_image_region}
\end{figure}

To estimate the possible diffuse-X-ray emissions 
around the bow shock region numerically, 
we first defined three regions: `BD43', `BSK', and `BGD', 
representing BD+43$^\circ$3654, its bow shock, and the background 
areas, respectively. 
The definitions of these areas are shown in Fig.\ \ref{fig:xray_image_region}. 
The circular areas around the XMM sources are excluded 
from the BSK and BGD regions; 
the radii of these excluded regions 
were defined by their luminosities.
The geometrical areas of BD43, BSK, and BGD were 
11.5, 41.2, and 44.8 arcmin$^2$, respectively.
Then, we counted X-ray events detected with the XIS, 
taking into account vignetting effects of the XRTs, 
we obtained $288.1 \pm 11.9$, $81.6 \pm 2.6$, 
and $74.0 \pm 2.3$ cnt arcmin$^{-2}$, 
for BD43, BSK, and BGD, respectively, 
where the errors are the 99\% statistical ones.
Thus, the events in the BSK region exceed those in the BGD region
by $7.6 \pm 3.4$ cnt arcmin$^{-2}$ statistically.

The measurement of the enhanced counts in the BSK region 
( $7.6 \pm 3.4$ cnt arcmin$^{-2}$) has 
systematic errors in estimations of the NXB and CXB. 
That of NXB is reported as 5\% by \citet{xis_bgd08}, 
and we adopt this value in this paper.
The systematic error in the CXB level is caused by 
fluctuations in numbers of unresolved point sources 
(which are mainly active galactic nuclei).
According to the systematic studies of CXB with ASCA by \citet{cxb02}, 
the CXB flux within the field of view of the GIS 
(Gas Imaging Spectrometer; \cite{Ohashi_96_gis1}; 0.5 deg$^2$)
fluctuated by 6.5\% in the 2 to 10 keV band.
Since the size of the BSK region is about 42 arcmin$^2$, 
the fluctuation would be 43\% for the BSK region by the XIS.
As described in session \ref{section:ana_image},
we removed dimmer point sources from the XIS data
than the GIS observation; i.e., the flux limit for the GIS was
$S_{\rm c} = 2 \times 10^{-13}$ erg s$^{-1}$ cm$^{-2}$ \citep{cxb02},
whereas our analyses achieved $4 \times 10^{-14}$ erg s$^{-1}$ cm$^{-2}$
(section \ref{section:ana_image}).
If we adopt the index of the $\log N$ -- $\log S$ relation 
for the population of point sources in CXB is 1.5, 
the CXB flux in our observation should have 12\% fluctuation
after removal of point sources with XMM-Newton.
Therefore, the systematic errors of NXB and CXB are 
5\% and 12\% for this FOV, respectively.
Numerically, the systematic errors for the NXB and CXB 
in the count rates of BD43, BSK, and BGD 
were $\pm 1.89 \pm 20.3$, $\pm 1.91 \pm 19.2$, 
and $\pm 5.69 \pm 27.8$ cnt arcmin$^{-2}$, respectively.
Thus, the enhancement was $7.6 \pm 3.4 \pm 6.0 \pm 34$ cnt arcmin$^{-2}$,
where the first, second, third errors represent 
statistics, systematics in the NXB estimation, and 
systematics in the CXB estimation, respectively; 
i.e., the X-ray count rate in the BSK region 
is consistent with that in the BGD region 
within the systematic errors.

The X-ray spectrum of the BSK region, 
in which the vignetting-corrected CXB and the NXB were subtracted, 
can be well reproduced by a power-law model 
with a photon index of $1.10^{+0.37+0.18}_{-0.33-0.03}$ 
and an X-ray flux of $ 2.56 ^{+0.83+0.70}_{-0.74-0.59}\times 10^{-13}$ 
erg s$^{-1}$ cm$^{-2}$ in the 0.5--10 keV band, 
where the first and second errors, respectively, correspond to 
the statistical error and the combined systematic errors for CXB and NXB, 
at the 3 sigma level. 
Thus, the 3-sigma upper limit of the X-ray luminosity from the BSK region
in the 0.5--10 keV band is 
$1.1 \times 10^{32}$ erg s$^{-1}$ at a distance of 1.4 kpc.

\section{Discussion}
\label{section:discussion}
We searched for diffuse X-ray emissions from the shock region
of BD+43$^\circ$3654 with Suzaku (see Section \ref{section:obs}),
and found an upper limit for the diffuse X-ray luminosity of 
$1.1 \times 10^{32}$ erg s$^{-1}$ per 41.2 arcmin$^2$
in the 0.5--10 keV band (see Section \ref{section:ana_flux_diffuse}).
By analogy with SNR systems
(see Section \ref{section:introduction:SNRcomparison}),
if the termination shock occurs in the region, 
interstellar materials would be shock-heated, 
and electrons or protons would be shock-accelerated. 
We discuss the efficiency of the heating by shock 
in Section \ref{section:discussion_thermal}, 
and the acceleration process 
in Section \ref{section:discussion_nonthermal}.

\subsection{Efficiency of thermalization by the shock}
\label{section:discussion_thermal}
If we assume that the value of $v_{\rm sw}$ for the object is the typical value 
of 2300 km~s$^{-1}$ \citep{Markova04,Repolust04} and that
the stellar gas has a typical temperature of $T_{\rm is} \sim 10^5$ K and 
acts as an ideal gas having a heating ratio of $\gamma = \frac{5}{3}$, 
then the sound speed in the gas is
$c_{\rm s} = 4 \times 10^6 \left(\frac{T_{\rm is}}{10^5 {\rm K}}\right)^{0.5}
\left(\frac{\gamma}{5/3}\right)^{0.5}$ cm  s$^{-1}$.
This value is much lower than $v_{\rm sw}$.
Therefore, we can expect a strong termination shock
to occur inside the contact discontinuity facing the runaway star,
while the bow shock occurs on the opposite side.
In this situation, the cold materials of the stellar wind 
could be shock-heated to a temperature of
\begin{eqnarray}
kT^{\rm sh} &=& 
\frac{3}{16} \frac{\mu m_{\rm H}}{k}
\left(\frac{4v_{\rm sw}}{3}\right)^2 \nonumber \\
&=& 11.2 \left(\frac{\mu}{0.615}\right)
\left(\frac{v_{\rm sw}}{2300 ~{\rm km~s}^{-1}}\right)^2 {\rm keV},
\label{eq:shock_temperature}
\end{eqnarray}
where $\mu$, $m_{\rm H}$, and $k$ are 
the mean ratio of numbers of electrons and nucleons,
the mass of hydrogen, and Boltzmann constant, respectively.
In other words, we can expect hot plasmas to exist 
immediately behind the termination shock.

In the rest frame of the star, 
the interstellar-matter (ISM) gas flows toward the star
and forms shocks through interaction with the stellar wind.
The incoming energy of the ISM flow per unit time is
$(\rho_{\rm ISM}v_{\rm 0}^3/2)\times(4\pi R_0{}^2)$, 
where we have neglected enthalpy, and $v_{\rm 0}$ is 
the stellar bulk speed through the ISM.
This gives a smaller contribution to the energy source of a
possible thermal emission than the ejected kinetic
luminosity of the stellar wind,
$\dot{E}_{\rm sw}\sim (1/2)\dot{M}v_{\rm sw}^2
\sim 2\pi\rho v_{\rm sw}^3R_0^2$, 
by a factor of $F=v_{\rm sw}/v_{\rm 0}\sim40$,
assuming $v_{\rm 0} \sim v_\star$.
Here, $\rho_{\rm ISM}$ and $\rho$ are the densities of uniform ISM 
and the stellar wind just in front of the termination shock, respectively,
and we simply assume the pressure balance,
$\rho_{\rm ISM}v_{\rm 0}^2\sim\rho v_{\rm sw}^2$, 
and a spherical shape for the termination shock with radius $R_0$.
However, the shocked, hot region is compressed to have 
a density $F^2\rho\sim1.6\times10^3\rho$ at most,
and thus we can expect an anisotropy of the emission from hot
plasmas: it should be brighter toward the runway direction as in the radio
synchrotron image \citep{Benaglia10}.

The Suzaku observation represents the upper limit of the X-ray luminosity 
at $< 1.1 \times 10^{32}$ erg s$^{-1}$, which corresponds to 
the emission measure EM of $< 3.8 \times 10^{54}$ cm$^{-3}$ for plasmas 
with the temperature of $kT = 11.2$ keV 
as expected by equation (\ref{eq:shock_temperature}). 
This value of the upper limit of EM reduces only by factor 4 
even if we change $kT$ by one order of magnitude lower.
The EM can be described as
$
{\rm EM} = \int n_{\rm e,th}^2 dV = \bar{n}_{\rm e,th}^2 V_{\rm BS} \eta,
$
where $n_{\rm e,th}$ is the density of thermal electrons, 
$V$ is the volume of the plasma, 
$\bar{n}_{\rm e,th}$ is the averaged value of electron density, 
$V_{\rm BS}$ is the volume of the bow shock region 
($1.5 \times 10^{56}$ cm$^3$; \cite{Benaglia10}), and 
$\eta$ is a density-weighted filling factor of the plasma 
in the bow shock region.
If we adopt the density of cold electrons 
$n_{\rm e,cl} \sim 6$ cm$^{-3}$ \citep{Comeron07}, 
the upper limit of EM can be written into the following constraint;
\begin{equation}
\left(\frac{\bar{n}_{\rm e,th}}{ n_{\rm e,cl}}\right)^2 \eta 
< 0.7 \times 10^{-3}.
\label{equation:thermal_limit}
\end{equation} 
Therefore, the shock-heating process is inefficient in the shock
region of the runaway star BD+43$^\circ$3654.
If all the cold materials are heated into the hot plasma 
($n_{\rm e,th} = n_{\rm e,cl}$), for example, then 
a very small fraction of the shock region would be shock-heated; 
i.e., when a top thin region emits thermal X-rays 
as discussed in \citet{delValle12},
the thickness of the region would be less than $2 \times 10^{15}$ cm.  
Otherwise, if the all the shock region emit thermal X-rays ($\eta = 1$), 
then only $\left(\frac{\bar{n}_{\rm e,th}}{ n_{\rm e,cl}}\right) \sim$ 3 \% 
of cold electrons are heated by shocks. 
We expect further deep survey by a micro-calorimeter onboard 
the near future X-ray mission, ASTRO-H.

\subsection{Constraints on a possible particle acceleration 
process by stellar winds}
\label{section:discussion_nonthermal}
\begin{figure}[hbt]
\centerline{\includegraphics[width=8cm]{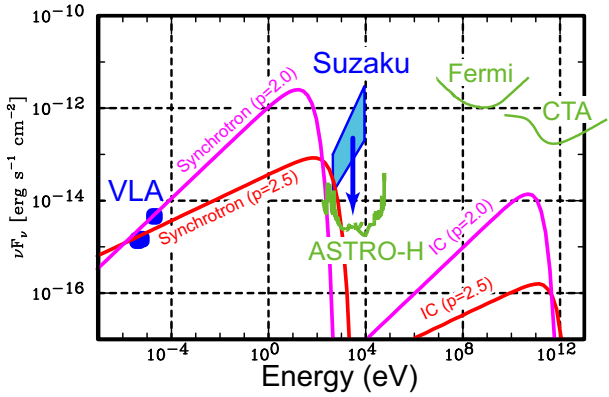}}
\caption{
The wide band energy spectrum of the shocked region of 
BD+43$^\circ$3654 from the radio to the TeV band.
The observation data from VLA and Suzaku are plotted in blue.
The red and magenta lines show the energy distributions of synchrotron 
and inverse-Compton emissions from electrons with 
a Lorentz factor of 1to $2.0 \times 10^7$ or 1 to $1.0 \times 10^7$,
and energy index $p$ = 2.5 or 2.0, respectively.
In the calculation, we assume $u_{\rm e} = u_{\rm B} \sim$ 
$6.9 \times 10^{-11}$ erg/cm$^3$ or $2.7 \times 10^{-11}$ erg/cm$^3$,
($B \sim$ 42~$\mu$G or 25~$\mu$G) for $p$ = 2.5 or 2.0, respectively.
CXB is only the source of inverse Compton emission in this calculation. 
The green lines show the sensitivities of 
the future X-ray mission {\it ASTRO-H} 
(\cite{astroh10} and references therein),
the Cherenkov Telescope Array (CTA) North, 
and the gamma-ray satellite {\it Fermi}.}
\label{fig:discussion_sed}
\end{figure}

As indicated by \citet{Benaglia10}, 
high-energy electrons with $p \sim 2.0$ on average
should exist around the shocked region 
and emit synchrotron radio radiation. 
From the X-ray observations obtained with Suzaku, 
we obtained a 3-sigma upper-limit of the X-ray luminosity 
at $1.1 \times 10^{32}$ erg s$^{-1}$ in the 0.5 to 10 keV band.
Fig.\ \ref{fig:discussion_sed} shows the multi-wavelength spectra 
from the shocked region of BD+43$^\circ$3654.
We have also plotted the energy spectra of synchrotron emissions 
for $p =$ 2.5 or 2.0,
assuming that the synchrotron emitting region is a sphere 
with a radius of $3 \times 10^{18}$ cm and 
that the energy densities of electron and magnetic fields 
are in equipartition (i.e., $u_{\rm e} = u_{\rm B}$).
The model calculation code is adopted from \citet{tashiro09}.
In the calculation, we assume 
$u_{\rm e} = u_{\rm B} = 6.9 \times 10^{-11}$ erg/cm$^3$ 
or $2.7 \times 10^{-11}$ erg/cm$^3$, 
corresponding to magnetic field strengths of 
$B \sim$ 42~$\mu$G or 25~$\mu$G 
for $p =$ 2.5 or 2.0, respectively.
Therefore, to account for the radio and X-ray luminosities,
the maximum Lorentz factor for electrons, $\gamma_{\rm max}$, should be
less than $10^7$, when $p$ is larger than $2.0$.
In other words,
the maximum energy of accelerated electrons 
would be $E_{\rm max} \leq 10 {\rm ~or~} 5~{\rm TeV}$
for $p =$ 2.5 or 2.0, respectively,
corresponding to a roll-off frequency of 
$\nu_{\rm roll} \leq 10^{17}~{\rm Hz}$.

Under the diffusive-shock-acceleration (DSA) theory \citep{Bell78}, 
particles need to be scattered back to the shock front multiple times,
so magnetic-field turbulence in the upstream region of the shock 
is crucial for the DSA mechanism.
Let us consider the time scales for the constraint of the magnetic-field 
turbulence, which is in the Bohm limit for PWNe and SNRs
(c.f., \cite{shibata03,bamba03,Bamba05}).
The cooling time scale of electrons with the maximum energy 
via synchrotron emission
($\sim 2.0$~kyr and 4.1~kyr for $p=2.5$ and 2.0, respectively,
from equation~(\ref{eq:timescale_sync}))
is longer than the dynamical timescale of $\sim 1$~kyr
(see equation~(\ref{eq:timescale_dy})).
Thus $E_{\rm max}$ is determined from the balance
of the acceleration and the dynamical timescales,
and we obtain
\begin{eqnarray}
E_{\rm max} &=& \frac{3}{20}\frac{1}{\xi}
\frac{v_s{}^2}{c}eB\tau_{\rm dy} \nonumber\\
 &=& \frac{60}{\xi}
\left(\frac{v_s}{\rm 2300~km~s^{-1}}\right)^2 \nonumber\\
&& \times
\left(\frac{B}{25~\mu{\rm G}}\right)
\left(\frac{\tau_{\rm dy}}{3 \times 10^{10}~{\rm s}}\right)\ {\rm TeV},
\label{eq:magnetic_turbulence}
\end{eqnarray}
where $\xi \sim (B/\delta B)^2$ is the ratio of the mean free path
and the gyroradius of electrons
\citep{skilling75,jokipii87,bamba03,nakamura10}.
In the Bohm limit, $\xi$ becomes 1.
Thus, the maximum energy should be $\sim 60/\xi$ or $108/\xi$~TeV,
for p=2.5 ($B\sim42~\mu$G) and p=2.0 ($B\sim25~\mu$G),
respectively.
Observationally, these values should be less than 10 and 5~TeV, 
respectively, and so, in either case, 
we find that $\xi$ should be larger than $\sim$ 11.
Therefore, the magnetic field might not be as turbulent
as in PWNe and SNRs.

Non thermal emission from protons, as well as electrons, 
could contribute possible Gamma-ray emission in Fermi and
CTA band \citep{Benaglia10,delValle12}.
The low level of turbulence from our results in the X-ray band
with the assumption of equipartition between $u_e$ and $u_m$
indicates low acceleration efficiency, 
not only for electrons but also for protons.
In this situation, the maximum energy of protons must be far below
the knee energy of $10^{15}$~eV. 
Furthermore, if the magnetic-field turbulence is generated by accelerated
protons themselves, as in the case of SNRs \citep{lucek2000},
the fact that $\xi>11$ suggests a lower density of cosmic-ray protons
in this system, which implies the proton injection rate is smaller 
than that in SNRs.

\section*{Acknowledgements}
The authors would like to thank all the members of the Suzaku team 
for their continuous contributions in the maintenance of onboard instruments, spacecraft operation, calibrations, software development,
and user support both in Japan and the United States. 
We would also like to thank the referee Dr. Andrei Bykov 
and the journal editor for careful readings and useful comments.
This work was supported in part
by Grants-in-Aid for Scientific Research (B) from 
the Ministry of Education, Culture, Sports, Science and Technology (MEXT)
(No.~23340055, Y.~T; No.~22340039, M.S.~T),
a Grant-in-Aid for Young Scientists (A) from MEXT
(No.~22684012, A.~B.),
a Grant-in-Aid for Young Scientists (B) from MEXT
(No.~21740184, R.~Y.),
and a Grant-in-Aid for Research Fellowship for Young Scientists (DC2)
from the Japan Society for the Promotion of Science 
(No.~239311, T.~K.).
Finally, YT and AB would like to thank Saki Terada for continuous supports.


\end{document}